\documentclass[amsmath,aps,prb,twocolumn,showpacs]{revtex4}

\usepackage{graphicx}
\usepackage[utf8]{inputenc}
\usepackage[T1]{fontenc}

\DeclareGraphicsExtensions{.eps}

\newcommand{\etal}[2][]{\emph{et~al.}#1\cite{#2}}
\newcommand{\wn}{\ensuremath{~\text{cm}^{-1}}}
\newcommand{\veps}{\varepsilon}
\newcommand{\tsub}[1]{$_{#1}$}

\begin{document}
\title{Optical phonons, spin correlations, and spin-phonon coupling in the frustrated pyrochlore magnets CdCr\tsub2O\tsub4 and ZnCr\tsub2O\tsub4}
\author{Ch.~Kant}
\author{J.~Deisenhofer}
\email{joachim.deisenhofer@physik.uni-augsburg.de}
\author{T.~Rudolf}
\author{F.~Mayr}
\author{F.~Schrettle}
\affiliation{Experimental Physics~V, Center for Electronic Correlations and Magnetism, University of Augsburg, D-86135~Augsburg, Germany}
\author{V.~Gnezdilov}
\affiliation{B.I. Verkin Institute for Low Temperature Physics and Engineering, NASU, UA-61103~Kharkov, Ukraine}
\author{D.~Wulferding}
\author{P.~Lemmens}
\affiliation{Institute for Physics of Condensed Matter, Technical University of Braunschweig, D-38106~Braunschweig, Germany}
\author{A.~Loidl}
\affiliation{Experimental Physics~V, Center for Electronic Correlations and Magnetism, University of Augsburg, D-86135~Augsburg, Germany}
\author{V.~Tsurkan}
\affiliation{Experimental Physics~V, Center for Electronic Correlations and Magnetism, University of Augsburg, D-86135~Augsburg, Germany} 
\affiliation{Institute of Applied Physics, Academy of Sciences of Moldova, MD-2028~Chişinău, Republic of Moldova}

\date{\today}

\begin{abstract}
We report on infrared, Raman, magnetic susceptibility, and specific heat measurements on CdCr\tsub2O\tsub4 and ZnCr\tsub2O\tsub4 single crystals. We estimate the nearest-neighbor and next-nearest neighbor exchange constants from the magnetic susceptibility and extract the spin-spin correlation functions obtained from the magnetic susceptibility and the magnetic contribution to the specific heat. By comparing with the frequency shift of the infrared optical phonons above $T_N$, we derive estimates for the spin-phonon coupling constants in these systems. The observation of phonon modes which are both Raman and infrared active suggest the loss of inversion symmetry below the Néel temperature in CdCr\tsub2O\tsub4 in agreement with theoretical predictions by Chern and coworkers [Phys.~Rev.~\textbf{B} 74, 060405 (2006)]. In ZnCr\tsub2O\tsub4 several new modes appear below $T_N$, but no phonon modes could be detected which are both Raman and infrared active indicating the conservation of inversion symmetry in the low-temperature phase.
\end{abstract}


\maketitle

\section{Introduction}

\begin{figure}[b]
\includegraphics[width=0.35\textwidth]{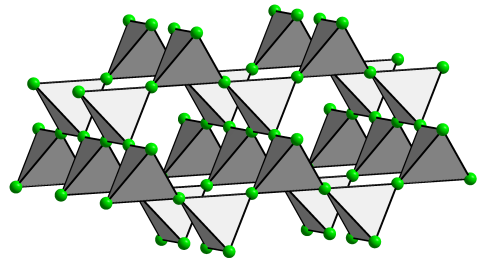}
\caption{\label{fig:SpinelStructure}(Color online) The pyrochlore
structure formed by the magnetic Cr ions on the $B$-site in
$A$Cr\tsub2$X_4$ ($X$ = S, O) spinels.}
\end{figure}

The antiferromagnetic  oxide-spinel systems $A$Cr\tsub2O\tsub4 ($A$~= Cd, Mg, Zn) are  prototypical examples for highly frustrated magnets,\cite{lee02} where the magnetic Cr$ ^{3+}$ ions with spin $s=3/2$ reside on the vertices of corner-sharing tetrahedra forming a pyrochlore lattice as shown in Fig.~\ref{fig:SpinelStructure}. An antiferromagnetic nearest-neighbor (nn) Heisenberg exchange on the pyrochlore lattice leads to inherent frustration and a multiply degenerate magnetic ground state. In real systems this degeneracy will be released at finite temperatures by coupling to other degrees of freedom and, indeed, the $A$Cr$_2$O$_4$ systems ($A$ = Cd, Mg, Zn) undergo a magnetostructural transition with antiferromagnetic order at 7.8, 12.7, and 12.5~K, \cite{rovers02, lee00} although their Curie-Weiss temperatures ($\Theta_{CW}$) are $-71$, $-346$, and $-390$~K, respectively.\cite{rudolf07a,sushkov05} The theory of the magnetoelastic coupling and the magnetostructural transition for these systems was worked out by Tchernyshyov and coworkers in terms of a spin-driven Jahn-Teller effect.\cite{tchernys02}

The structural and magnetic  ground state properties in the Cr spinels are strongly intertwined and, hence, the assignment of the correct symmetry groups is a complex task which has not been accomplished yet: The high-temperature cubic structure of spinels has space group $Fd\bar{3}m$, which is often reduced to tetragonal with space group $I4_1/amd$ \cite{akimitsu78,hidaka03,radaelli05,reehuis03} when undergoing a structural or magnetostructural transition. Such a uniform tetragonal distortion was reported for CdCr\tsub2O\tsub4.\cite{chung05,lee07} However, it was suggested from a theoretical analysis of the observed spiral spin structure that the lattice distortion should be chiral with space group $I4_122$, leading to the loss of inversion symmetry (IS).\cite{chern06} The corresponding distortion in ZnCr\tsub2O\tsub4 has been mostly described in terms of the tetragonal $I\bar{4}m2$ space group also lacking inversion symmetry,\cite{lee00,lee07,ji09} but orthorhombic distortions have also been evoked leading to $F222$ symmetry\cite{glazkov09} (no IS) or $Fddd$ symmetry (conserving IS).\cite{kagomiya02}

The number of infrared (IR) and Raman active optical phonons strongly depends on the exact symmetry and occupied Wyckoff positions of the lattice. If inversion symmetry is present IR and Raman modes are mutually exclusive. Hence, the appearance of resonances which are IR and Raman active is a secure indication for the loss of inversion symmetry.\cite{poulet76} While we are not aware of any low-temperature Raman studies of these systems, the magnetoelastic coupling and the phase transitions clearly show up in the far-infrared optical spectra as shifts or splittings of the phonon modes.\cite{sushkov05,rudolf07a,aguilar08,rudolf09} For ZnCr\tsub2O\tsub4 the spin-phonon coupling was estimated via two different approaches: The first one is a direct evaluation of the phonon splitting in the magnetically ordered phase using the spin-driven Jahn-Teller effect and the second involves a comparison of the frequency shift of the phonons with the spin-spin correlation function derived from the magnetic contribution to the specific heat.\cite{sushkov05} The first approach also has been applied to CdCr\tsub2O\tsub4 already.\cite{aguilar08} 

In this study we investigate  the low-temperature Raman and IR optical response for both systems and analyze in detail the approach to derive the spin-phonon coupling by considering both, the spin-spin correlation function derived from the magnetic contribution to the specific heat, and the one obtained directly from the paramagnetic susceptibility. The outline of the paper is as follows: In section \ref{sec:magnsuscep} we describe the susceptibilities of CdCr\tsub2O\tsub4 and ZnCr\tsub2O\tsub4 by a Quantum Tetrahedral Meanfield model \cite{garcia-a00} and obtain the nn and next-nearest neighbor (nnn) exchange couplings in these systems. Then we extract the magnetic part of the specific heat in Sec.~\ref{sec:specheat}. After discussing the phonon excitations observed by IR and Raman measurements with regard to the proposed lattice symmetries in section \ref{sec:phonon}, we estimate in Sec.~\ref{sec:ad} the spin-phonon coupling in both compounds by comparing the frequency shift of the IR active phonons to the spin-spin correlations obtained from the susceptibility and the specific heat data.

\section{Experimental details}

High-quality platelike single crystals of CdCr\tsub2O\tsub4 up to 7~mm in diameter were obtained by spontaneous crystallization from Bi\tsub2O\tsub3-V\tsub2O\tsub5 flux. ZnCr\tsub2O\tsub4 single crystals were grown by a chemical transport technique using Cl as a transport agent at temperatures between 950 and 900ºC. The length of the octahedra edges extents up to 6~mm. The optical measurements were performed on as grown (111) mirror-like surfaces. X-ray powder diffraction patterns of crashed single crystals did not reveal any impurity phases and correspond to stoichiometric polycrystals. The magnetic susceptibility measurements were carried out in a commercial superconducting quantum interference design magnetometer (Quantum Design MPMS-5). The heat capacity was measured in a Quantum Design Physical Properties Measurement System for temperatures from 2~K < $T$ < 300~K. The IR reflectivity was determined using a Bruker Fourier-Transform IR spectrometer IFS 113v/S, which was equipped with a He-bath and a He-flow cryostat (5~K< $T$ < 300~K). Raman scattering experiments were carried  out in a quasi-backscattering geometry. An Ar/Kr ion laser was used for excitation at 514.5~nm (2.41~eV). The laser output power was kept below 3~mW on a focus of approximately 50~$\mu$m of diameter to protect the sample from possible heating effects. The scattered light was collected and dispersed by a triple monochromator DILOR XY on a liquid-nitrogen-cooled CCD detector in parallel and crossed polarization configurations. Temperature dependencies were obtained in a variable temperature closed cycle cryostat (Oxford/Cryomech Optistat, RT-2.8 K).

\begin{figure}[t]
\includegraphics[width=0.4\textwidth]{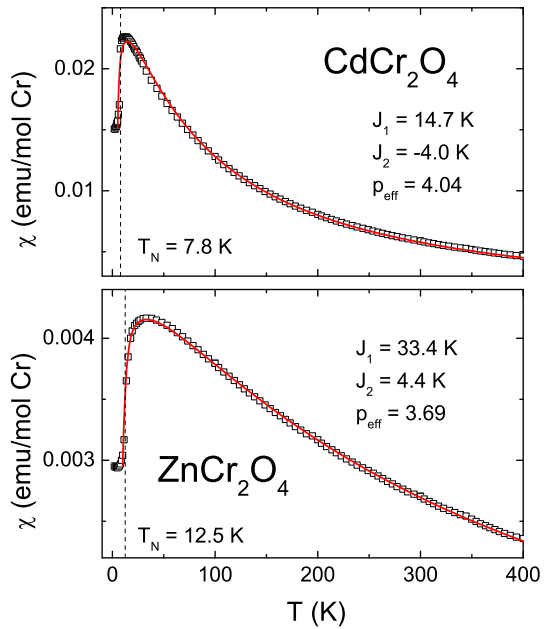}
\caption{\label{fig:CCOchi}(Color online) Temperature dependence of the magnetic susceptibility of CdCr\tsub2O\tsub4 and ZnCr\tsub2O\tsub4 in a field of $\mu_0H=1$~T and $\mu_0H=0.1$~T, respectively. The solid lines are fits to the experimental data according to Eq.~(\ref{eq:chiTMF}).}
\end{figure}

The measured reflectivity spectra are analyzed by modeling the dielectric  function and fitting it to the experimental data using the program ReFFit.\cite{kuzmenko05} We utilize a three-parameter model for each phonon mode which is a sum over Lorentzian oscillators:
\begin{equation}\label{eq:LModel}
\varepsilon(\omega)=\sum_j \frac{\Omega_j}{\omega_{0j}^2-\omega^2-i\gamma_j \omega} + \veps_\infty
\end{equation}
Here, $\Omega_j$ denotes the plasma frequency, $\omega_{0j}$ the eigenfrequency, and $\gamma_j$ the damping of mode $j$, respectively. $\veps_\infty$ is the dielectric constant beyond the phonons in the mid-infrared regime and accounts for high-frequency electronic polarizations. It is treated as a free parameter in the fitting procedure. The optical phonons in CdCr$_2$O$_4$ and ZnCr$_2$O$_4$ were measured from 5 to 300~K in the far-infrared and additional measurements were performed in the mid-infrared up to 8000\wn{} for CdCr\tsub 2O\tsub 4 to determine the value of $\veps_{\infty}$ = 4.6 with high accuracy and to enable a Kramers-Kronig transformation. Therefor a constant extrapolation of the measured reflectivity was assumed at low frequencies and a smooth $\omega^{-h}$ behavior at the upper limit of the recorded spectrum.

\section{Experimental Results and Discussion}

\subsection{Magnetic Susceptibility}\label{sec:magnsuscep}

Figure \ref{fig:CCOchi} shows the magnetic susceptibility  of CdCr$_2$O$_4$ and ZnCr\tsub 2O\tsub4 as a function of temperature. On cooling both curves exhibit a broad maximum before magnetic order sets in at the indicated Néel temperatures. Note that the susceptibility below $T_N$ was shown to depend strongly on the applied magnetic field.\cite{martinho01a} Our results are in good agreement with previous studies.\cite{rovers02,martinho01}

García-Adeva and Huber derived a Quantum Tetrahedral Meanfield (TMF) model\cite{garcia-a00} to calculate the magnetic susceptibility of pyrochlore lattices. We use this model in the form \begin{equation}\label{eq:chiTMF}
\chi_{TMF}(T) = \frac{N_A g^2 \mu_B^2}{k_B} \frac{a \cdot \chi_{tet}(T)}{1+(3 J_1 +12 J_2)\chi_{tet}(T)},
\end{equation}
which describes the susceptibility per mole of magnetic ions. Here, $N_A$ is the Avogadro constant,  $g=1.97$ the $g$-factor,\cite{martinho01} $\mu_B$ the Bohr magneton, and $J_1$ and $J_2$ are the nn and nnn exchange constants (in units of $k_B$), respectively. $\chi_{tet}$ can be calculated through
\begin{equation}
\chi_{tet}(T) = \frac{1}{12 T} \frac{\sum\limits_S g(S) S (S+1) (2S+1) e^{\frac{-J_1 S (S+1)}{2 T}}} {\sum\limits_S g(S) (2S+1) e^{\frac{-J_1 S (S+1)}{2 T}}}.
\end{equation}
The sum runs over the total spin values $S=(0,1,2,3,4,5,6)$ of the Cr-tetrahedron  and $g(S)=(4,9,11,10,6,3,1)$ are the corresponding degeneracies.\cite{garcia-a00} The scaling factor $a$ is related to the effective number of Bohr magnetons by $a=p^2_{eff} / [g^2 s(s+1)]$.

The solid lines in Fig.~\ref{fig:CCOchi} were  obtained by fitting Eq.~(\ref{eq:chiTMF}) to the   experimental data. This procedure yielded $J_1=14.7$~K, $J_2=-4.0$~K, and $p_{eff}=4.04$ for CdCr$_2$O$_4$ and $J_1=33.4$~K, $J_2=4.4$~K, and $p_{eff}=3.69$ for ZnCr$_2$O$_4$. In literature values for $p_{eff}$ range from 3.98 to 4.02~$\mu_B$ for the Cd spinel\cite{rovers02,rudolf07a} and between 3.85 and 3.94~$\mu_B$ for the Zn compound.\cite{leccabue93,rudolf07a} The spin only value for a Cr$^{3+}$ ion of $3.87\mu_B$ corresponds well to the reported data and to our findings. The exchange coupling constants were estimated from the Curie-Weiss temperatures to be $J_1=12$~K (Ref.~\onlinecite{chung05}) for the Cd and $J_1=35 - 45$~K (Ref.~\onlinecite{martinho01}) for the Zn compound which is also in good agreement to our results. In Ref.~\onlinecite{garcia-a02b} the TMF model was already applied to ZnCr\tsub2O\tsub4 and yielded $J_1=39.4$~K and $J_2=1.76$~K. Evidently, $J_1$ is about one order of magnitude stronger than $J_2$ in ZnCr\tsub2O\tsub, while in CdCr\tsub2O\tsub4 the antiferromagnetic $J_1$ is already weakened and the nnn exchange $J_2$ is ferromagnetic. This is in agreement with the reduction of the direct exchange between nn Cr ions with an increase of the lattice parameter when substituting Zn by Cd.\cite{rudolf07a}

We also attempted to analyze our susceptibility data with an alternative model proposed by García-Adeva and Huber,\cite{garcia-a02b,garcia-a01a,garcia-a01,garcia-a02} which, however, did not allow to fit the data satisfactorily.\cite{kant09}

\subsection{Specific Heat}\label{sec:specheat}

\begin{figure}
\includegraphics[width=0.43\textwidth]{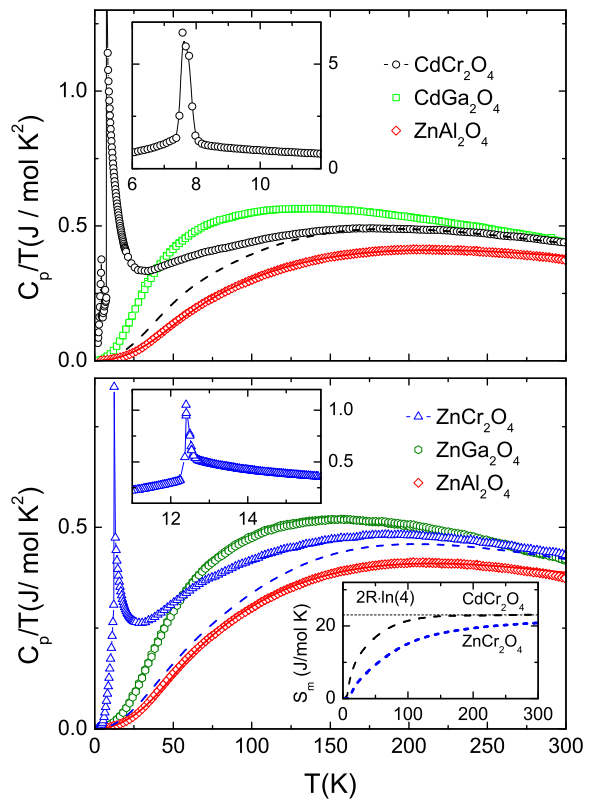}
\caption{\label{fig:ABOcp}(Color online) Upper frame: Temperature dependence of $C_p/T$ of CdCr\tsub2O\tsub4 (open circles), CdGa\tsub2O\tsub4 (open squares), and the reference compound ZnAl\tsub2O\tsub4 (open rhombi) together with the fitted lattice contributions $C^{fit}_{lat}/T$ of the specific heat (lines) as described in the text. Upper insets: $C_p/T$ as a function of temperature around the AFM transition. Lower frame: $C_p/T$ of ZnCr\tsub2O\tsub4 vs.\ temperature (open triangles) together with a fit of the phonon contribution (dashed line). Also shown are ZnGa\tsub2O\tsub4 (open hexagons) and the reference compound. Lower inset: Entropy of the magnetic contribution $C_p-C^{fit}_{lat}$ of the specific heat for both Cr spinels.}
\end{figure}

In Fig.~\ref{fig:ABOcp} we show the temperature dependence of specific heat divided by temperature $C_p/T$ for CdCr\tsub2O\tsub4, ZnCr\tsub2O\tsub4 and the non-magnetic compounds CdGa\tsub2O\tsub4, ZnGa\tsub2O\tsub4, and ZnAl\tsub2O\tsub4. The magneto-structural transitions of CdCr\tsub2O\tsub4 at $T_N=7.8$~K and ZnCr\tsub2O\tsub4 at $T_N=12.5$~K are clearly visible as very sharp peaks, indicating first-order phase transformations.

In order to extract the magnetic  contribution to the specific heat we followed the procedure used in Ref.~\onlinecite{martinho01} for ZnCr\tsub2O\tsub4: We prepared the non-magnetic reference compounds CdGa\tsub2O\tsub4 and ZnGa\tsub2O\tsub4 to subtract the lattice contribution to the specific heat directly. The obtained specific heat data are also displayed in Fig.~\ref{fig:ABOcp}. Obviously, the data for the reference compounds do not allow for a direct subtraction, and it is not easy to explain the discrepancies for ZnGa\tsub2O\tsub4 in comparison to Ref.~\onlinecite{martinho01}. Therefore, we chose an alternative approach to extract the magnetic contribution. As a non-magnetic reference we prepared ZnAl\tsub2O\tsub4 and fitted the specific heat data by modeling the phonon density of state (PDOS). We were able to describe the experimentally obtained values with one isotropic Debye- ($D$) and three isotropic Einstein-type ($E_{1,2,3}$) contributions satisfactorily (compare rhombi and solid line in Fig.~\ref{fig:ABOcp}) with the corresponding Debye and the Einstein temperatures $\theta_D$ and $\theta_{E_{1,2,3}}$, respectively. To comply with the 21 degrees of freedoms per formula unit, we fixed the ratio between the Debye- and Einstein-terms to $D:E_1:E_2:E_3=1:1:3:2$. The parameters obtained by a best fit are $\theta_D=330$~K, $\theta_{E_1}=350$~K, $\theta_{E_2}=640$~K, and $\theta_{E_3}=1080$~K for ZnCr\tsub2O\tsub4. The derived Einstein temperatures are in good agreement with maxima in the PDOS reported by neutron diffraction studies.\cite{fang02}

Turning now to the frustrated systems ZnCr\tsub2O\tsub4 and CdCr\tsub2O\tsub4 we modeled the phonon contribution by scaling the Debye temperature of our reference with $\sqrt{m_\textrm{ZAO}/m_{A\textrm{CO}}}$ as it is expected for the phonon dispersion of acoustic modes close to the center of the first Brillouin zone. $m_\textrm{ZAO}$ denotes the molar mass of the reference and $m_{A\textrm{CO}}$ of the Cr spinel, respectively. The Einstein terms were adapted in such a way, that the obtained magnetic contribution $C_m=C_p-C^{fit}_{lat}$ results in a magnetic entropy (see lower inset of Fig.~\ref{fig:ABOcp})
\begin{equation}
S_m(T)= \int_{0}^T\frac{C_m}{\vartheta}\textrm d\vartheta,
\end{equation}
which increases continuously with temperature and approaches the expected high-temperature limit value of $2R\cdot \ln(2s+1)$ for a spin system with $s=3/2$. For CdCr\tsub2O\tsub4 this procedure yielded $\theta_D=265$~K, $\theta_{E_1}=300$~K, $\theta_{E_2}=570$~K, and $\theta_{E_3}=820$~K. In ZnCr\tsub2O\tsub4 we obtained $\theta_D=290$~K, $\theta_{E_1}=360$~K, $\theta_{E_2}=600$~K, and $\theta_{E_3}=860$~K.

\subsection{Infrared and Raman modes}\label{sec:phonon}

\subsubsection{Infrared spectroscopy}

\begin{figure}
\includegraphics[width=0.45\textwidth]{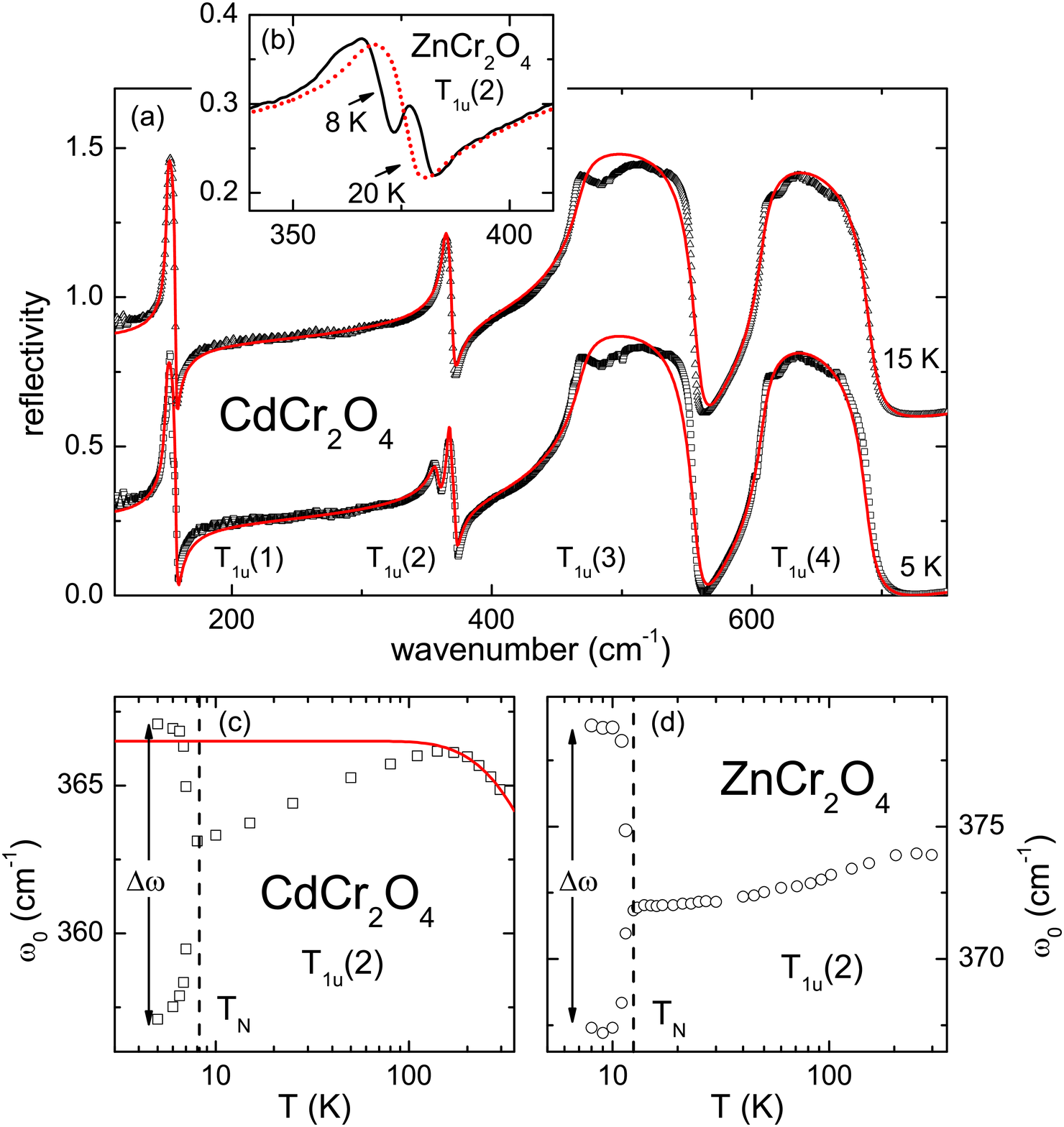}
\caption{\label{fig:CCOphonon}(Color online) (a) Reflectivity spectra of CdCr\tsub 2O\tsub 4  in  the paramagnetic state at 15~K (shifted for clarity) and at 5~K, below $T_N$. The $T_{1u}$(2) mode is clearly split at 5~K.\cite{rudolf07a} (b) Splitting of the $T_{1u}$(2) phonon in ZnCr\tsub 2O\tsub 4. (c), (d) Temperature dependence of the phonon eigenfrequencies of the $T_{1u}$(2) modes revealing the splitting at $T_N$ in CdCr\tsub 2O\tsub 4 and ZnCr\tsub 2O\tsub 4, respectively.}
\end{figure}

In Fig.~\ref{fig:CCOphonon}(a) we show reflectivity  spectra for CdCr$_2$O$_4$ at 5~K and 15~K, below and above the magnetostructural transition at $T_N=7.8$~K. The fits using Eq.~(\ref{eq:LModel}) are shown as solid lines in the same plot. The four IR active phonon modes $T_{1u}(j)$ (with $j=1, 2, 3, 4$) observed above $T_N$ (150, 365, 465, and 605\wn{} at room temperature), were labeled according to the triply degenerate modes expected from the irreducible representations in the cubic $Fd\bar{3}m$ (\#227) symmetry: \cite{rousseau81}
\begin{align*}
\Gamma = & \; 4 T_{1u}                              & \text{(IR active)}\\
         & +  A_{1g} + E_{1g} + 3 T_{2g}            & \text{(Raman active)}\\
         & + 2 A_{2u} + 2 E_{u} + T_{1g} + 2 T_{2u} & \text{(silent)}
\end{align*}

For ZnCr\tsub2O\tsub4 the reflectivity spectra exhibit a qualitatively similar shape and have been published earlier.\cite{sushkov05} The four reststrahlen bands are located at room temperature at  186, 373, 509, and 620\wn{}, respectively. Therefore, we focus on the mode $T_{1u}(2)$ at 8~K and 20~K, below and above the Néel temperature $T_N=12.5$~K in Fig.~\ref{fig:CCOphonon}(b). As previously reported,\cite{sushkov05,rudolf07a,aguilar08} this mode softens upon cooling and splits for both compounds when entering the magnetically ordered phase [see Figs.~\ref{fig:CCOphonon}(c) and (d)]. 
The solid line in Fig.~\ref{fig:CCOphonon}(c) is a fit of the increase of the phonon frequency with decreasing temperature using
\begin{equation} \label{eq:anharmfreq}
\omega_j(T) = \omega_{0j}\cdot \left \lbrack 1- \frac{c_j}{\exp(hc\; \omega_{av}/ k_B T)-1}\right \rbrack,
\end{equation}
to include anharmonic effects.\cite{wakamura88a} Here, $\omega_{0j}$ indicates the eigenfrequency of the phonon in the absence of spin-phonon coupling at 0~K, $c_j$ is a mode dependent scaling factor of the anharmonic contributions, and $\omega_{av}=423\wn{}$ is the arithmetic average of the IR- and Raman-active phonon frequencies at room temperature.  

\begin{figure}
\includegraphics[width=0.45\textwidth]{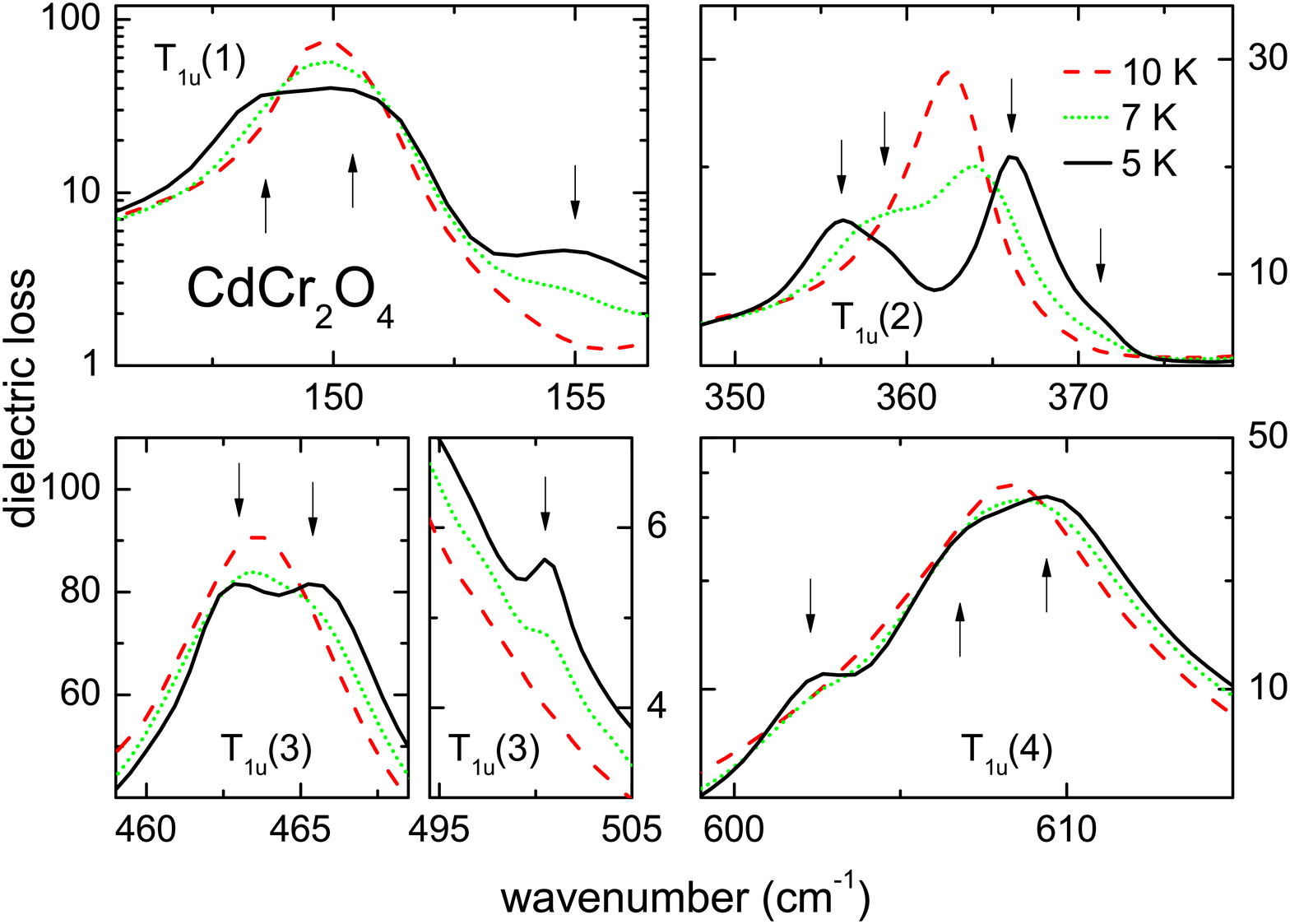}
\caption{\label{fig:CCOe2}(Color online) Dielectric loss of the phonons in CdCr$_2$O$_4$. Below $T_N$ one can clearly see the splitting of the cubic modes and the appearance of new modes (indicated by arrows).}
\end{figure}

In these AFM Cr spinels magnetic ordering is accompanied by  a structural symmetry reduction. Accordingly, a splitting of the degenerate cubic $T_{1u}(j)$ modes and the appearance of new modes is expected. The most common structural distortion for spinels occurring below the Néel temperature is tetragonal with space group $I4_1/amd$ (\#141) \cite{akimitsu78,hidaka03,radaelli05,reehuis03} with the following irreducible representations:\cite{rousseau81}
\begin{align*}
\Gamma = & \; 4A_{2u} + 6E_{u}                    & \text{(IR active)}\\
         & + 2A_{1g} + 3B_{1g} + B_{2g} + 4E_g    & \text{(Raman active)}\\
         & + 2A_{1u} + A_{2g} + 2B_{1u} + 4B_{2u} & \text{(silent)}
\end{align*}
Such a symmetry reduction reportedly takes place in CdCr$_2$O$_4$, where the incommensurate spin structure is stabilized by a tetragonal distortion with an elongated $c$-axis.\cite{chung05,lee07} Hence, one expects a splitting of each of the four cubic modes into a singlet and a doublet and the appearance of two more doublet modes. When zooming in on the dielectric loss spectra in the region of the four modes in CdCr$_2$O$_4$ in Fig.~\ref{fig:CCOe2}, splittings of all modes can be detected upon entering into the ordered state similar to the previously reported case of polycrystalline samples.\cite{rudolf07a}

Note that the effects are very subtle for the $T_{1u}(1)$ and $T_{1u}(4)$ modes and, therefore, we chose a semi-logarithmic plot for these phonons. The $T_{1u}(1)$ mode develops into a broad flat feature indicative of a two-peak structure (at 148.6 and 150.4\wn{}) and at about 155\wn{} a new mode emerges. The splitting of the second phonon $T_{1u}(2)$ into two modes at 256.2 and 366.1\wn{}, which has been described earlier in Ref.~\onlinecite{rudolf07a}, is strong compared to the other phonons and amounts to approximately 10\wn{}. Note that both split modes exhibit a shoulder on the high-frequency side which can be interpreted as two more new modes at 358.7 and 371.3\wn{}. The phonon $T_{1u}(3)$ features a rather clear splitting into two modes at 463 and 465.4\wn{} and a weak new mode appears at 500.5\wn{}. On the low-energy side of the $T_{1u}(4)$ mode, at about 602.3\wn{}, an additional peak becomes visible below $T_N$ and the main resonance at 609.4\wn{} exhibits a shoulder at 606.8\wn{}. Therefore, we identify 13 IR active phonon modes below $T_N$ for CdCr$_2$O$_4$. Since from the factor group analysis of space group $I4_1/amd$ only ten IR phonons should be observable, we conclude that the true space group should be of lower symmetry than $I4_1/amd$.\cite{chung05,lee07}

In ZnCr$_2$O$_4$ the reported commensurate  spin structure is accompanied by a tetragonal distortion with a contracted $c$-axis and the space group $I\bar{4}m2$ (\#119) has been assigned.\cite{lee00,lee07} The corresponding factor-group analysis for a $1\times 1\times 1$ chemical unit cell for ZnCr$_2$O$_4$ yields the following symmetry distribution of $\Gamma$-point vibrational modes:

\begin{align*}
\Gamma =\; &8 B_2 + 11 E &(\text{IR and Raman active})\\
           &+ 6 A_1+ 3 B_1 &(\textrm{Raman active})\\
           &+3 A_2   & (\textrm{silent})\\
\end{align*}

\begin{figure}[b]
\includegraphics[width=0.4\textwidth]{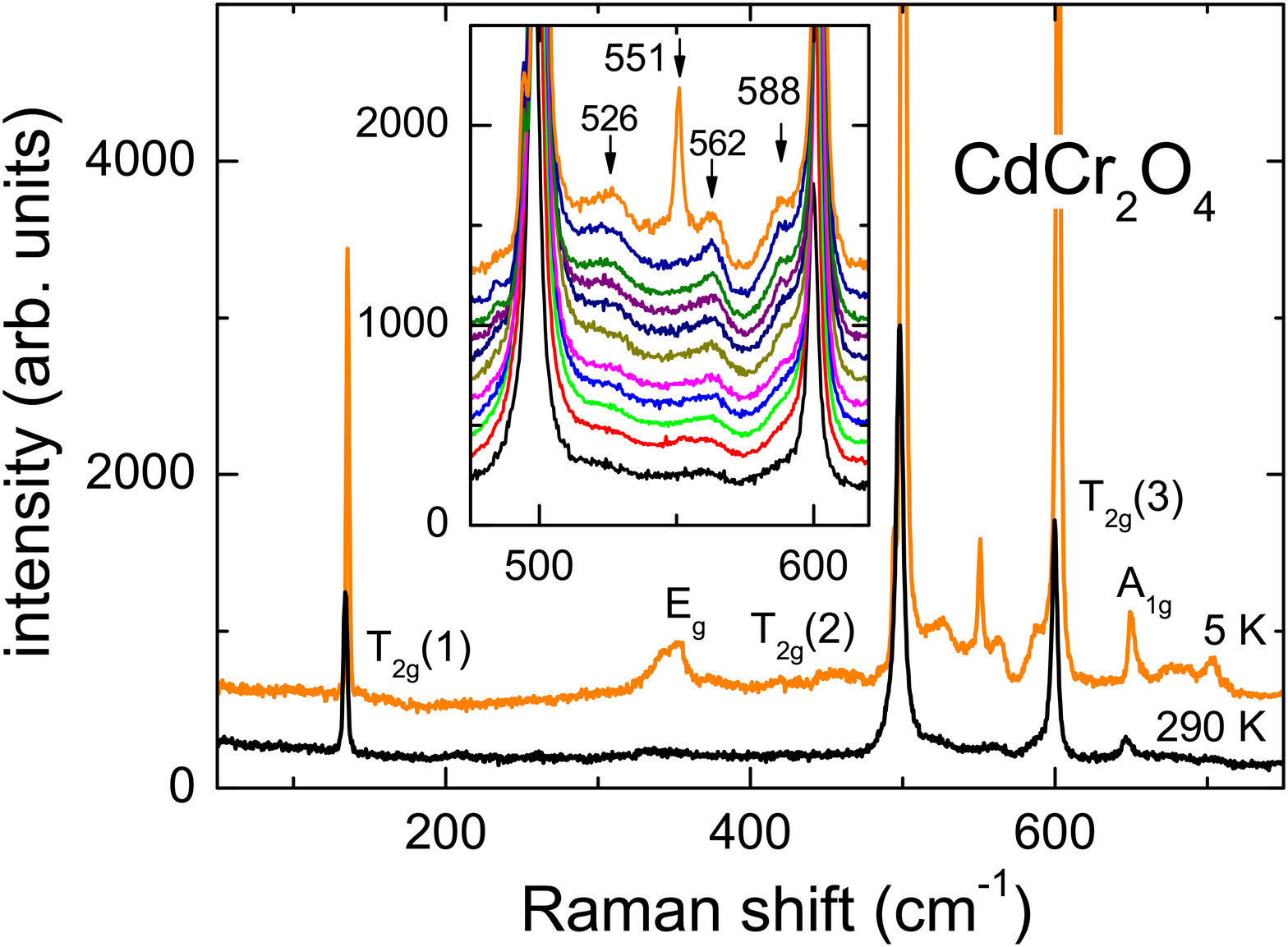}
\caption{\label{fig:ccoRa1}(Color online) Raman spectra of  CdCr\tsub2O\tsub4 for various temperatures. The spectrum at 5~K is shifted for clarity. In the magnetically ordered state a sharp resonance appears at 551\wn{}. Inset: temperature evolution of Raman spectra in the region of the $T_{2g}(2)$ and $T_{2g}(3)$ modes ($T$= 5, 10, 50, 100, 150, 175, 200, 225, 250, 275, and 290~K from top to bottom). Additional features are indicated by arrows and their Raman shift is given in \wn{}. Spectra below 290~K are shifted for clarity.}
\end{figure}

Note, that in this symmetry the $8 B_2$ and $11 E$ modes are both Raman and IR active, because of the lack of inversion symmetry. Recently, ZnCr\tsub2O\tsub4 was studied using synchrotron x-rays and neutron diffraction.\cite{ji09} To determine the tetragonal structure in detail, the x-ray integrated intensity of about 140 different superlattice reflections was measured and fitted within $I\bar{4}m2$ symmetry. Cr ions were found to occupy six crystallographically distinct sites: four $8i$ sites and two $16j$ sites. Note that expanding the tetragonal unit cell by $\sqrt{2}\times \sqrt{2} \times 2$ compared to the cubic (or simply $c$-axis contracted tetragonal) unit cell should lead to the appearance of even more extra lines in the Raman and IR spectra.

A x-ray powder diffraction study was also performed for ZnCr\tsub2O\tsub4 by Kagomiya \etal{kagomiya02a} and a Rietveld refinement below $T_N$ was performed for the orthorhombic $Fddd$ symmetry (\#70). For this structure the factor group analysis gives:

\begin{align*}
\Gamma =\; &7 B_{1u} + 7 B_{2u} + 7 B_{3u} &(\textrm{IR active})\\
  &+3 A_g + 4 B_{1g} + 4 B_{2g} + 4 B_{3g}&(\textrm{Raman active})\\ 
  &+6A_u &(\textrm{silent}) \\
\end{align*}

Here one would expect 21 IR-active and 15 Raman-active phonons. In principle,  the number of observed phonons in the IR and Raman data could distinguish between the proposed space groups. In the IR experiment, however, only the splitting of $T_{1u}(2)$ and a new mode at 553\wn{} has been observed,\cite{sushkov05,rudolf07a} and no final conclusion can be drawn from IR data alone. Therefore, we will now turn to the Raman spectra of CdCr\tsub2O\tsub4 and ZnCr\tsub2O\tsub4.

\subsubsection{Raman scattering}

\begin{figure}[b]
\includegraphics[width=0.32\textwidth]{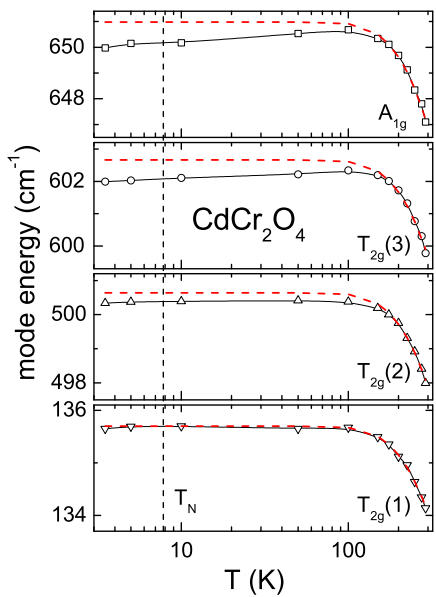}
\caption{\label{fig:ccoRa2}(Color online) Temperature evolution of the four cubic Raman modes.  The dashed lines are fits using Eq.~(\ref{eq:anharmfreq}). Solid lines are drawn to guide the eye.}
\end{figure}

The Raman spectra for CdCr\tsub2O\tsub4 at 290~K and at 5~K, below $T_N = 7.8$~K, are shown in Fig.~\ref{fig:ccoRa1}. The five expected Raman modes for the cubic structure can be identified in the high-temperature data at 290~K. The spectrum consists of three phonon triplets of $T_{2g}$ symmetry (134, 499, and 600\wn{}), one weak doublet of $E_g$ symmetry (343\wn{}), and one singlet of $A_{1g}$ symmetry (647\wn{}).

The temperature dependence of the eigenfrequencies of the four intensive Raman-active modes of CdCr\tsub2O\tsub4 are shown in Fig.~\ref{fig:ccoRa2}. With decreasing temperature, the phonon eigenfrequencies increase, as usually observed in anharmonic crystals. The data for $T > 150$~K were fitted using Eq.~(\ref{eq:anharmfreq}) with  $\omega_{av}$ = 423\wn{} (dashed lines in Fig.~\ref{fig:ccoRa2}). The $T_{2g}(1)$ mode shows a purely anharmonic behavior while three phonon modes, $T_{2g}(2)$, $T_{2g}(3)$, and $A_{1g}$, reveal negative deviations for $T < 150$~K. We observe no abrupt shifts of the cubic Raman modes at $T_N$.

\begin{figure}[t]
\includegraphics[width=0.25\textwidth]{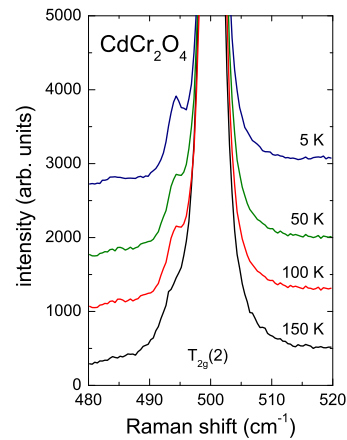}
\caption{\label{fig:ccoRa3}(Color online) Temperature evolution of the $T_{2g}(2)$ mode and its split-off resonance at the low-frequency side (spectra are shifted for clarity).}
\end{figure}

Based on its energy and on the fact that it has the highest intensity among the other very weak lines in the Raman spectra, the mode at about 334~\wn{} is identified as the cubic $E_g$ phonon mode. This mode is about four times as broad as the $T_{2g}$ and $A_g$ phonon lines. Besides, a shoulder is identified on its high energy side, which has the same frequency vs.\ temperature behavior as the main line. However, it grows faster in intensity upon cooling. Higher order multiphonon scattering is observed in the Raman spectra, too. We identified lines with energies of approximately 670 and 700\wn{} (not shown here) as overtones of the $E_g$ mode and its satellite.

In the tetragonal $I4_1/amd$ phase at temperatures below $T_N$ one expects a splitting of the triplet modes into a doublet and a singlet and the appearance of two more singlet modes, overall ten Raman active modes. However, additional weak and broad features (indicated by arrows in the inset of Fig.~\ref{fig:ccoRa1}) appear in the region of the $T_{2g}(2)$ and $T_{2g}(3)$ modes already at temperatures far above $T_N$. Only the feature at 551\wn{} seems to be directly related to the low-temperature phase, because it is still absent at 10~K, just above the magnetostructural transition. When zooming in on the Raman spectra in the region of the $T_{2g}(2)$ mode in Fig.~\ref{fig:ccoRa3}, a weak split-off resonance at 493\wn{} on the low-frequency side is clearly seen upon cooling, similar to the feature at 588\wn{} at the low-frequency side of $T_{2g}(3)$.

At present it is not clear where the additional modes observable above $T_N$ stem from, but such effects may be attributed to a modification of the selection rules similar to observations in resonance Raman scattering in semiconductors.\cite{kauschke87} They can be separated into violations of long-wavelength selection rules as purely intrinsic bulk effects and additional extrinsic or defect contributions. Three main processes and their origin have been identified:\cite{martin71,Martin75} (i) q-dependent, intraband Fröhlich interactions, (ii) q-independent, impurity-induced effects, and (iii) surface electric field effects at a depletion layer. All three of the given mechanisms might be relevant in the present case, because comparably large intensities are observed and defects cannot be ruled out completely. Further studies including a test of polarization selection rules on samples with different defect contributions under resonant and nonresonant conditions would be needed to further clarify the origin of the additional modes.

Despite the unknown origin of these additional resonances, the Raman data provides valuable information on the low-temperature symmetry properties of CdCr\tsub2O\tsub4, because the low-temperature eigenfrequencies of the $T_{2g}(2)$ and $T_{2g}(3)$ modes are identical with two of the weak new modes appearing in the IR data below $T_N$. This strongly suggests that below $T_N$ inversion symmetry is lost in agreement with the theoretical predictions by Chern and coworkers, whose analysis of the spiral magnetic ground state favored a chiral structure with space group $I4_122$.\cite{chern06}

\begin{figure}
\includegraphics[width=0.4\textwidth]{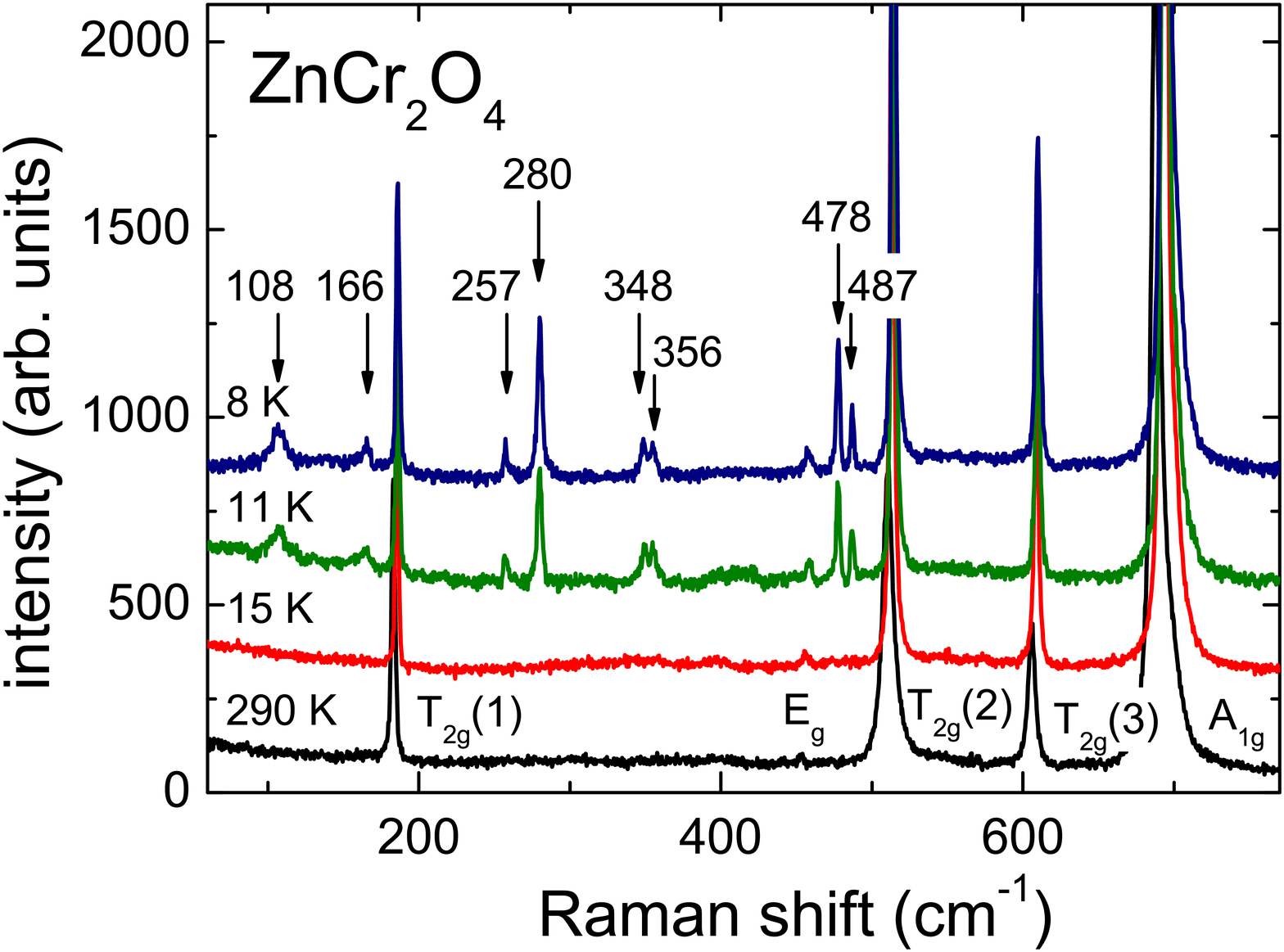}
\caption{\label{fig:raman1}(Color online) Raman spectra of ZnCr\tsub2O\tsub4 at selected temperatures measured at parallel polarization configuration. Spectra below room temperature are shifted for clarity by a constant offset. Arrows indicate new phonon lines at temperatures below $T_N$.}
\end{figure}

We now discuss the Raman spectra of ZnCr\tsub2O\tsub4 depicted in Fig.~\ref{fig:raman1} for selected temperatures below and above the magnetostructural transition at $T_N$ = 12.5~K. Our measurements at room temperature agree with previously published data for single crystals of ZnCr\tsub2O\tsub4.\cite{lutz91,himmrich91} As in CdCr\tsub2O\tsub4 we observe the three triplets of $T_{2g}$ symmetry (183, 511, and 606\wn{}), the weak doublet of the $E_g$ symmetry (457\wn{}), and the singlet of $A_{1g}$ symmetry (688\wn{}).

\begin{figure}[b]
\includegraphics[width=0.3\textwidth]{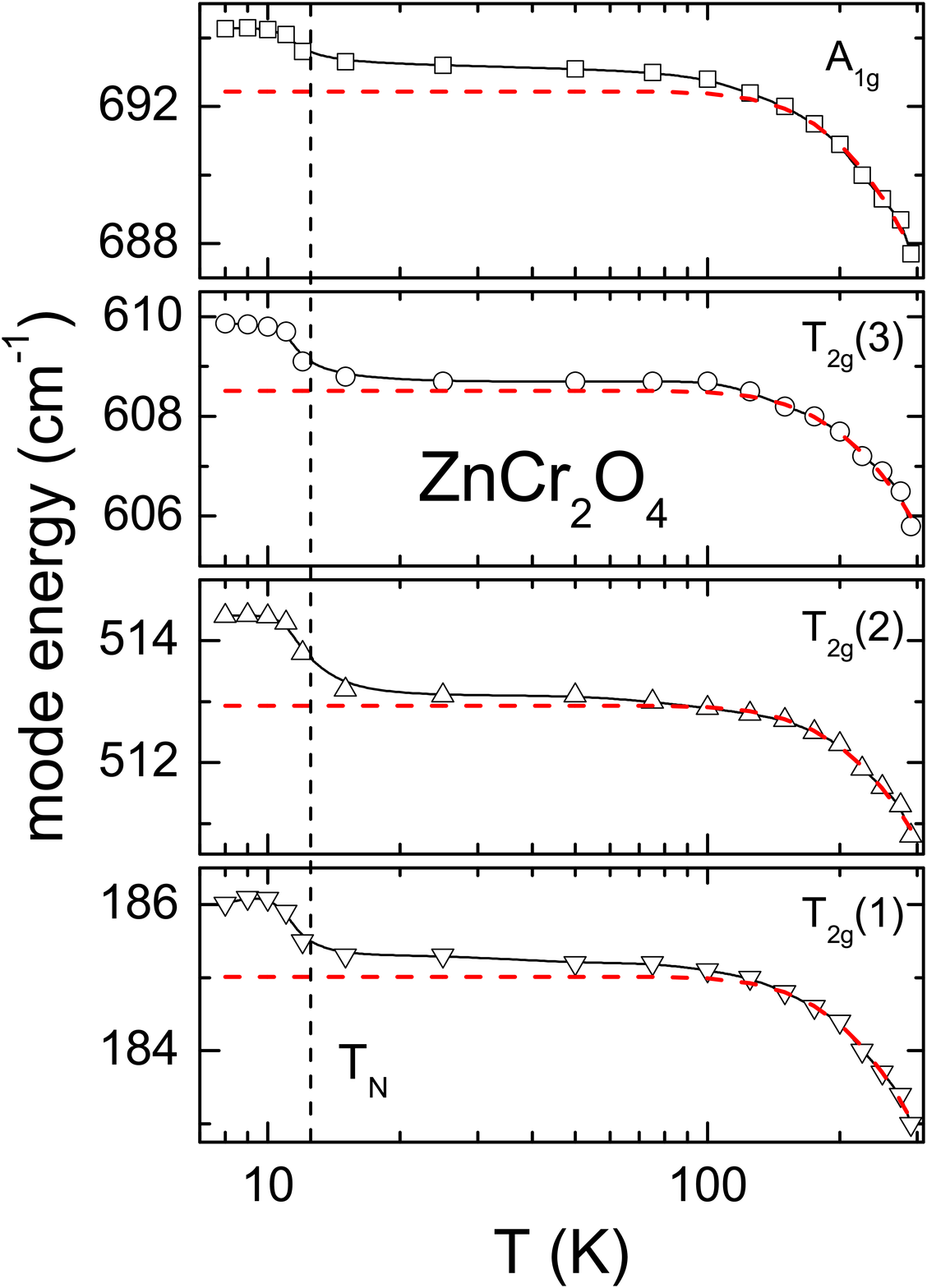}
\caption{\label{fig:raman2}(Color online) Mode energy of the phonons in ZnCr\tsub2O\tsub4 as a function of temperature. The dashed lines are fits to the high-temperature results using a simple anharmonic term, Eq.~(\ref{eq:anharmfreq}).}
\end{figure}

\begin{figure}[t]
\includegraphics[width=0.42\textwidth]{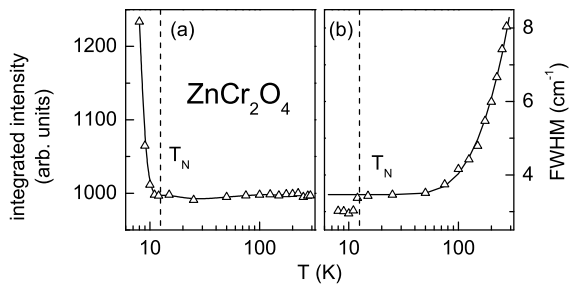}
\caption{\label{fig:raman3}Temperature dependence of the integrated intensity (a) and the line width (FWHM) (b) of the $T_{2g}$ (510\wn{}) triplet mode. The solid line in (a) is draw in to guide the eye and reflects in (b) the expected behavior for ZnCr\tsub2O\tsub4 according to Eq.~(\ref{eq:anharmDamp}).}
\end{figure}

The temperature dependence of the eigenfrequencies of the four intensive Raman-active modes of ZnCr\tsub2O\tsub4 are shown in Fig.~\ref{fig:raman2}. The dashed lines are fits using Eq.~(\ref{eq:anharmfreq}). Evidently, all modes reveal deviations from this purely anharmonic behavior for $T$<150~K. The discrepancies are of the order of 10$^{-3}$ and positive for all Raman-active modes. The anomalous temperature dependence smoothly evolves below 150~K, but significantly becomes enhanced just below $T_N$, where a jump of the eigenfrequencies occurs. We observed also an anomalous increasing of the phonon line intensities below $T_N$ and an abrupt breakdown of their width at $T_N$. Figures~\ref{fig:raman3}(a) and (b) show the $T$ dependence of the integrated intensity and line-width of the $T_{2g}(2)$ triplet mode, respectively.

Similar anomalies in the intensity were observed in related spinels early on and, subsequently, it has been shown that this resonant effect is connected with a change of the electronic zone structure caused by magnetic ordering.\cite{gunthero89} For example in CdCr\tsub2S\tsub4, resonant Raman scattering is related to a red shift in the absorption spectrum caused by the magnetic ordering.\cite{koshizuk76}

The solid line in Fig.~\ref{fig:raman3}(b) is a fit of the temperature dependence of the phonon line width using
\begin{equation}\label{eq:anharmDamp}
\Gamma(T) = \Gamma_0 \left \lbrack 1+\frac{2}{\exp(hc \; \omega_0/2 k_B T)-1} \right\rbrack
\end{equation}
with $\Gamma_0 = 3.46\wn{}$ and the eigenfrequency $\omega_0 = 516\wn{}$. The data can be described very well down to the Néel temperature by this expression, which incorporates cubic anharmonicity effects.\cite{klemens66,balkansk83}

Experimentally, besides the shifts of the five cubic phonon modes, we observe an appearance of eight new phonon modes with frequencies 106.7, 164.8, 257.3, 280, 348.4, 355, 477.6, and 487.0\wn{} in the Raman spectra of ZnCr\tsub2O\tsub4 below $T_N$ (indicated by arrows in Fig.~\ref{fig:raman1}). Unfortunately, the identification of a total of 13 Raman and six IR active modes does not allow to conclusively discard any of the proposed low-temperature symmetries. However, we would like to point out that, in contrast to CdCr\tsub2O\tsub4, we could not identify any phonon mode that appears in both Raman and IR spectra below $T_N$ as implied by the assignment of the $I\bar{4}m2$ symmetry \cite{lee00,lee07,ji09} or the additional orthorhombic lattice distortions within space group $F222$ (\#22) evoked to describe the antiferromagnetic resonance properties of ZnCr\tsub2O\tsub4.\cite{glazkov09}  Therefore, it seems very likely that at low-temperatures inversion symmetry is conserved and IR and Raman modes are mutually exclusive. This would single out the assignment of the orthorhombic $Fddd$ symmetry by Kagomiya \etal{kagomiya02a} as the possibly true space group in the magnetically ordered state.

\subsection{Spin-phonon coupling}\label{sec:ad}

In the Cr spinels spin-phonon coupling is considered to be the driving force to relieve magnetic frustration by a the magnetostructural phase transition. Already above the transition, in the correlated paramagnetic state, the spin-phonon coupling shows up in the temperature dependence of the phonon frequencies. In the following we will concentrate on the IR active $T_{1u}$(2) modes of CdCr\tsub2O\tsub4 and ZnCr\tsub2O\tsub4, because this mode exhibits a clear splitting upon entering into the magnetically ordered state [ Fig.~\ref{fig:CCOphonon}(c) and (d)]. With respect to the behavior expected for the frequency shift due to anharmonicity [solid line in Fig.~\ref{fig:CCOphonon}(c)], there is a clear deviation which becomes stronger on approaching magnetic ordering. Therefore, it has been attributed to the coupling of the lattice to spin fluctuations. Considering a nn Heisenberg spin system, Baltensberger and Helman related the frequency shift of lattice vibrations in magnetic crystals to the magnetic energy of the system,
\begin{equation}\label{eq:DwSS}
\Delta\omega \approx \lambda \langle s_i\cdot s_j\rangle,
\end{equation}
by introducing the spin-phonon coupling parameter $\lambda$, which depends on the derivatives of the exchange constants with respect to the coordinates of the magnetic ions.\cite{baltensp68,baltensp70} The coupling parameter $\lambda$ in the paramagnetic state can be determined if both the frequency shift of the lattice vibrations and the spin-spin correlation function are known. To obtain the spin-spin correlation function we follow two independent approaches.

According to the fluctuation-dissipation theorem the magnetic molar susceptibility per spin in the disordered state can be written as
\begin{equation}\label{eq:chiFDT}
\chi(T)=\frac{N_A g^2 \mu_B^2}{k_B T}\sum\limits_{m,n} \langle s_m \cdot s_n \rangle_\chi,
\end{equation}
where $\langle s_m \cdot s_n \rangle$ represents the spin-spin correlation function between the spin $s_m$ and $s_n$. Restricting the sum to nn, the spin-spin correlation function is given by \begin{equation}\label{eq:corrChi}
\langle s_i \cdot s_j \rangle_\chi=\frac{k_B T \chi(T)}{N_A g^2 \mu_B^2}-\frac{s (s+1)}{3};
\end{equation}
i.e., we can directly utilize the experimental susceptibility data shown in Fig.~\ref{fig:CCOchi}.

\begin{figure}[b]
\includegraphics[width=0.48\textwidth]{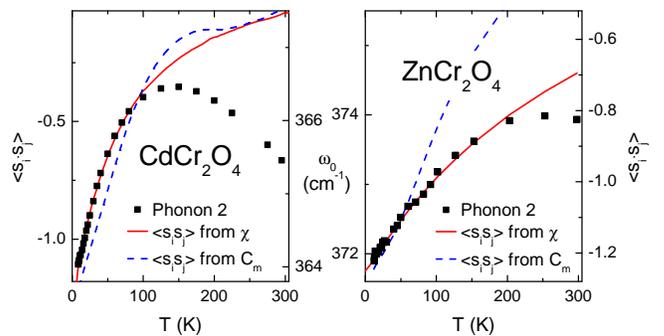}
\caption{\label{CCOsst}(Color online) Temperature dependence of the spin correlations $\langle s_i\cdot s_j\rangle$ derived according to Eqs.~(\ref{eq:corrChi}) and (\ref{eq:corrCm}) and the eigenfrequency $\omega_0$ of phonon $T_{1u}(2)$ in a scaling that reveals a linear behavior of the two quantities at low temperatures.}
\end{figure}

An alternative derivation of the nn spin-spin correlation function can be obtained from specific heat data.\cite{sushkov05} Assuming that the dominant part of the magnetic energy is given by the nn Heisenberg interaction, one can extract the temperature dependence of the spin-spin correlations from the magnetic contribution of the specific heat via
\begin{equation}\label{eq:corrCm}
\langle s_i\cdot s_j\rangle_{C_m}(T)=const + \frac{1}{6N_A J_1} \int_{T_N}^T
C_m(\vartheta)\textrm d\vartheta,
\end{equation}
where $6N_A$ is the number of bonds between adjacent magnetic ions per mole. For the analysis of the specific heat, we used the nn exchange coupling $J_1$ obtained from the magnetic susceptibility data derived in Sec.~\ref{sec:magnsuscep}.

In Fig.~\ref{CCOsst} we plot $\langle s_i s_j \rangle(T)$ calculated from  the magnetic susceptibility (solid line) and from the magnetic contribution of the specific heat (dashed line). In addition, the eigenfrequencies $\omega_0$ of the $T_{1u}$(2) mode are shown as a function of temperature (solid squares). The offsets and scalings of the ordinates were chosen to make $\omega_0$ coincide with the susceptibility data. Comparing the curves for the correlation functions of CdCr\tsub 2O\tsub 4, one can see that they show a similar evolution and a good agreement of their absolute values. At low temperatures both exhibit a quasi-linear behavior. With increasing temperatures they increase monotonically and reach zero at about room temperature. Up to 125~K the phonon frequency scales very well with the evolution of $\langle s_i s_j \rangle$. The similarity between the correlation curves can be regarded as a justification of the procedure to describe and extract the lattice contribution to the specific heat in Sec.~\ref{sec:specheat}. Since, however, $\langle s_i s_j \rangle_{C_m}$ suffers from the uncertainty of modeling the PDOS and depends on the exact determination of the nn exchange coupling $J_1$, we think that the use of $\langle s_i s_j \rangle_\chi$ leads to a more precise determination of $\lambda$.

\begin{figure}[b]
\includegraphics[width=0.38\textwidth]{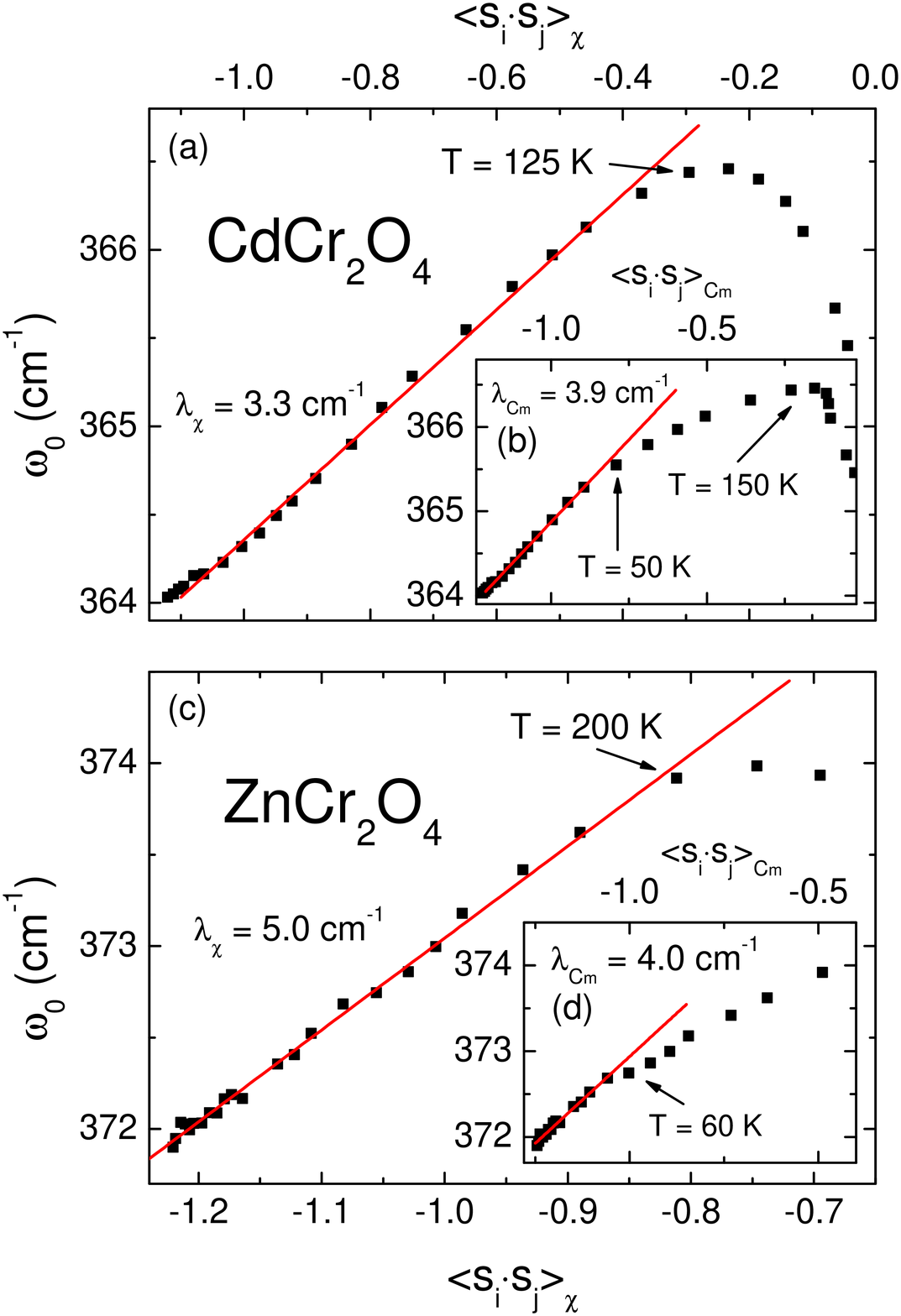}
\caption{\label{fig:wvsSS}(Color online) Eigenfrequency $\omega_0$ as function of the spin-spin correlation functions for CdCr\tsub2O\tsub4 and ZnCr\tsub2O\tsub4. (a), (c) $\omega_0$ vs.\ $\langle s_i s_j \rangle_\chi$ obtained from the magnetic susceptibility data. (b), (d) $\omega_0$ vs.\ $\langle s_i s_j \rangle_{C_m}$ derived from specific heat measurement.}
\end{figure}

Assuming that Eq.~(\ref{eq:DwSS}) holds, we directly plot the frequency shift $\Delta \omega$ vs.~the spin-spin correlation $\langle s_i s_j \rangle_\chi$ and $\langle s_i s_j \rangle_{C_m}$ for the measured temperatures above $T_N$ in Fig.~\ref{fig:wvsSS}. In Fig.~\ref{fig:wvsSS}(a) one clearly sees a linear behavior for CdCr\tsub 2O\tsub 4 up to about 100~K. The solid line is a linear fit to the data [Eq.~(\ref{eq:DwSS})] and yields a value for the spin-phonon coupling constant in the paramagnetic state of $\lambda_\chi = 3.3\wn{}$. At 125~K the eigenfrequencies begin to deviate from the regression curve. The inset [Fig.~\ref{fig:wvsSS}(b)] shows $\omega_0$ against $\langle s_i s_j \rangle_{C_m}$, the spin-spin correlation function calculated from the magnetic specific heat. Below 50~K the data can be fitted linearly giving a slope of $\lambda_{C_m} = 3.9\wn{}$. Above 50~K it starts to deviate until an abrupt breakdown above 150~K occurs. In the lower panels of Fig.~\ref{fig:wvsSS} we present our results for the ZnCr\tsub 2O\tsub 4 spinel. The plot of $\omega_0$ versus $\langle s_i s_j \rangle_\chi$ [Fig.~\ref{fig:wvsSS}(c)] shows a linear increase up to about 150~K. The slope gives a coupling constant of $\lambda_\chi = 5.0\wn{}$. When the frequency shift is plotted against $\langle s_i s_j \rangle_{C_m}$ we attain a value of 4.0\wn{}. This drawing reveals a convincing linear relation up to about 60~K and then starts to segregate from the regression curve [see Fig.~\ref{fig:wvsSS}(d)].

A comparable value $\lambda=6.2\wn{}$ has been reported for ZnCr\tsub2O\tsub4 by Sushkov \etal{sushkov05} using $\langle s_i s_j \rangle$ derived by subtracting the specific heat of non-magnetic ZnGa\tsub2O\tsub4 from ZnCr\tsub2O\tsub4. Although we could not follow their approach as described above, the agreement with our results $\lambda_\chi=5.0\wn{}$ and $\lambda_{C_m}=4.0$, is fairly good.

Keeping in mind that the spin-phonon coupling leads to the magnetostructural phase transition the splitting of the $T_{1u}$(2) modes allows to obtain further estimates for $\lambda$: The magnetostructural transition can be described by the spin-Peierls order parameter $n_{sp}=\langle S_1\cdot S_2- S_1\cdot S_4\rangle$. It accounts for the difference in tetrahedral bonds with ferromagnetic or antiferromagnetic coupling and vanishes accordingly in the undistorted paramagnetic state. The splitting between the two modes can then be expressed in terms of $n_{sp}$ (Ref.~\onlinecite{fennie06}):

\begin{equation} \label{eq:Dwsp}
\Delta \omega  \approx \lambda\langle S_1\cdot S_2- S_1\cdot S_4\rangle
\end{equation}

In the case of CdCr$_2$O$_4$, where the dominant distortion is a uniform elongation along the $c$-axis, the singlet mode is lower in energy and the order parameter becomes $n_{sp}=-9/4$ if one assumes a collinear ground state of the spins. For the proposed tetragonal distortion of ZnCr\tsub2O\tsub4 with space group $I\bar{4}m2$,\cite{lee07} $n_{sp}$ equals 9/2 when assuming an oversimplified model of  a collinear ground state.

Via this approach $\lambda=4.4\wn{}$ and $\lambda=2.6\wn{}$ have been extracted previously for CdCr\tsub2O\tsub4 and ZnCr\tsub2O\tsub4, respectively.\cite{sushkov05} We find that the value for CdCr\tsub2O\tsub4 compares nicely to our results derived in the paramagnetic phase. This indicates that the proposed order parameter based on a collinear spin state describes the magnetostructural transition correctly. Assuming that our value for $\lambda$ for ZnCr\tsub2O\tsub4 is valid to describe also the splitting below $T_N$, one can estimate $n_{sp}\approx 2$, confirming that the assumption of a collinear spin configuration does not hold for this compound.

\section{Summary}

Using a Quantum Tetrahedral Mean-Field model we fitted the susceptibilities  of CdCr\tsub2O\tsub4 and ZnCr\tsub2O\tsub4 and obtained the nn and nnn exchange coupling constants  $J_1=14.7$~K, $J_2=-4.0$~K, and $J_1=33.4$~K, $J_2=4.4$~K, respectively. Moreover, we analyzed the specific heat for these systems in comparison to nonmagnetic reference compounds. From both the susceptibility and the specific heat, we extracted the spin-spin correlation function in order to relate it to the shift of the IR phonon $T_{1u}(2)$ which shows the strongest splitting at the Néel temperature. We argue that the results derived by using directly the magnetic susceptibility appear more reliable, resulting in spin-phonon coupling parameters $\lambda = 3.3\wn{}$ and $\lambda = 5.0\wn{}$ for the Cd and Zn spinel, respectively. Additionally, we show that the observed IR and Raman spectra of CdCr\tsub2O\tsub4 provide evidence for a loss of inversion symmetry, thus ruling out the proposed $I4_1/amd$ symmetry in the magnetically ordered phase. For the low-temperature phase of ZnCr\tsub2O\tsub4 we depicted that the proposed symmetries $I\bar{4}m2$ and $F222$, which imply that phonon modes are both Raman and IR active, cannot be confirmed by our experiments and, hence, favor the orthorhombic $Fddd$ symmetry.

\begin{acknowledgments}
We thank A.~Krimmel, D.~L.~Huber, A.~J.~García-Adeva, Yu.~G.~Pashkevich, and O.~Tchernyshyov for fruitful discussions. This work was partly supported by the Deutsche Forschungsgemeinschaft DFG through the Collaborative Research Center SFB~484 (University of Augsburg) and the Project LE967/6-1 (Technical University of Braunschweig). V. G. acknowleges Ukrainian-Belorassian grant F29.1/014 for a partial support.
\end{acknowledgments}

\end{document}